\documentclass[runningheads]{llncs}

\usepackage{graphicx}
\usepackage[utf8]{inputenc}
\usepackage{multirow}
\usepackage{listings}
\usepackage{adjustbox}
\usepackage{hyperref}
\usepackage{paralist}
\usepackage{color}%
\usepackage{xcolor}%

\usepackage{atbegshi,picture}

\hypersetup{
        pdfauthor={Ali Parsai, Serge Demeyer, Seph De Busser},
        pdftitle={C++11/14 Mutation Operators Based on Common Fault Patterns},
        pdfkeywords={Software Testing, Mutation Testing, C++11/14, Mutation Operators},
    colorlinks,
    citecolor=black,
    filecolor=black,
    linkcolor=black,
    urlcolor=black
}

\begin{document}
\title{C++11/14 Mutation Operators \\Based on Common Fault Patterns}
\author{Ali Parsai\inst{1}\orcidID{0000-0001-8525-8198} \and
	Serge Demeyer\inst{1}\orcidID{0000-0002-4463-2945} \and
	Seph De Busser\inst{1}}
\authorrunning{A. Parsai et al.}

\institute{University of Antwerp, Middelheimlaan 1, 2020 Antwerp, Belgium
\email{\{ali.parsai,serge.demeyer\}@uantwerpen.be} 
\\ \email{seph.debusser@gmail.com}
}

\maketitle

\begin{abstract}

The C++11/14 standard offers a wealth of features aimed at helping programmers write better code. Unfortunately, some of these features may cause subtle programming faults, likely to go unnoticed during code reviews. In this paper we propose four new mutation operators for C++11/14 based on common fault patterns, which allow to verify whether a unit test suite is capable of testing against such faults. We validate the relevance of the proposed mutation operators by performing a case study on seven real-life software systems.


\end{abstract}

\section{Introduction}
\label{section:Introduction}
Nowadays, the process of software development relies more and more on automated software tests due to the developers interest in testing their software components early and often. %
The level of confidence in this process depends on the quality of the test suite. Therefore, measuring and improving the quality of the test suite has been an important subject in literature. Among many of the studied techniques, mutation testing is known to perform well for improving the quality of the test suite~\cite{Just2014}.  

The idea of mutation testing is to help identify software faults indirectly by improving the quality of the test suite through injecting an artificial fault (i.e. generating a \textit{mutant}) and executing the unit test suite to see whether the fault is detected~\cite{Papadakis2018}. If any of the tests fail, the mutant is said to detected, thus \textit{killed}. On the other hand, if all the tests pass, the test suite failed to detect the mutant, thus the mutant \textit{survived}. However, some mutants result in code which does not pass the compiler and these are called \textit{invalid mutants}. And in other situations, a mutant fails to change the output of a program for any given input hence can never be detected --- these are called \textit{equivalent mutants}.

A mutant is created by applying a transformation rule (i.e. \textit{mutation operator}) to the code that results in a syntactic change of the program~\cite{Jia2011}. Given an effective set of mutation operators, mutation testing can help developers identify the weaknesses in the test suite~\cite{Baker2013}. Nevertheless, designing effective mutation operators requires considerable knowledge about the coding idioms and the common programming faults often made in the language~\cite{Jia2011}. More importantly, good mutation operators should maximize the likelihood of \textit{valid} and \textit{non-equivalent} mutants~\cite{Delgado-Perez2015a}.

The first set of mutation operators were reported in King et al.~\cite{King1991}.  They were later implemented in  the tool Mothra which was designed to mutate the programming language FORTRAN77. With the advent of the object-oriented programming paradigm, new mutation operators were proposed to cope with specific programming faults therein~\cite{Kim2000}. This is a common trend in mutation testing: languages evolve to get new language constructs; some of these constructs cause subtle programming faults; after which new mutation operators get designed to shield against these common faults. For example, with the evolution of Java related languages, mutation operators have been designed to account for concurrent code~\cite{Bradbury2006}, aspect-oriented programming~\cite{Ferrari2008}, graphical user interfaces~\cite{Oliveira2015}, and Android applications~\cite{Deng2017}.

The C++11/14 standard (created in 2011 and 2014 respectively) offers a wealth of features aimed at helping programmers write better code~\cite{Stroustrup2014}. Most notably there is more type-safety and compile-time checking (e.g. static\_assert, override). Unfortunately, the standard also provides a few features that may cause subtle faults (e.g. lambda expressions, list initialization, \ldots). Our goal is to identify these sources of common faults and introduce new mutation operators that address them. While it is possible that some subset of these faults are addressed by C++99 mutation operators, previous experience shows targeted mutation operators prove useful in improving the test suite quality further~\cite{Chekam2017,Delgado-Perez2017}.

\vspace{1em}
\noindent
In this study, we seek to answer the following research questions:

\begin{compactitem}
\item \textbf{RQ1.} Which categories of C++11/14 faults are most likely to be made by programmers, and what are the corresponding mutation operators?
\item \textbf{RQ2.} To what extent do these mutation operators create valid, non-equivalent mutants?
\end{compactitem}

\vspace{1em}
\noindent
The rest of this paper is structured as follows:
In Section~\ref{section:Background} we provide the necessary background information about this study, and briefly discuss the related work.
In Section~\ref{section:Design} we discuss our approach to answering our research questions, and show our results in Section~\ref{section:Results}. 
Finally, we present our conclusions in Section~\ref{section:Conclusions} and highlight the future research directions rooted in this work.
\vspace*{-1em}
\section{Background and Related Work}
\label{section:Background}
In this section  we provide the necessary background information needed to comprehend the rest of the article and discuss the related work. First, we describe mutation testing and its related concepts. Then, we describe the new C++11/14 features, focusing on subtle faults that may be revealed via mutation testing.

\subsection{Mutation Testing}
Mutation testing is the process of inserting bugs into software(\textit{Mutants}) using a set of rules(\textit{Mutation Operators}) and then running the accompanying test suite for each inserted mutant. If all tests  pass, the mutant survived. If at least one test fails, the mutant is killed. If the mutant causes an error during compilation of the production code, it is invalid. A valid mutant that does not change the semantics of the program, thus making it impossible to detect, is called equivalent. %

An equivalent mutant is a mutant that does not change the semantics of the program, i.e. its output is the same as the original program for any possible input.
Therefore, no test case can differentiate between an equivalent mutant and the original program, which makes it undesirable. The detection of equivalent mutants is undecidable due to the halting problem~\cite{Offutt1997}. 
The only way to make sure there are no equivalent mutants in the mutant set is to manually inspect and remove all the equivalent mutants. However, this is impractical in practice. Therefore, the aim is to generate as few equivalent mutants as possible.

Mutation operators are the rules mutation testing tools use to inject syntactic changes into software. Most operators are defined as a transformation on a certain pattern found in the source code. The first set of mutation operators ever designed were reported in King et al.~\cite{King1991}. These mutation operators work on basic syntactic entities of the programming language such as arithmetic, logical, and relational operators. Offutt et al. came up with a selection of few mutation operators that are enough to produce high quality test suites  with a four-fold reduction of the number of mutants~\cite{Offutt1996}. Kim et al. extended the set of mutation operators for object-oriented programming constructs~\cite{Kim2000}.

Because of the complexity of parsing C++, building a mutation testing tool for C++ is almost equivalent to building a complete compiler~\cite{Irwin2001}. %
It is only with modern tooling, e.g. the Clang/LLVM compiler platform, that it became possible to write such tools without an internal parser.

Kusano et al. developed CCmutator, a mutation tool for multi-threaded C/C++ programs that mutates usages of POSIX threads and the C++11 concurrency constructs, but works on LLVM's intermediate representation instead of directly on C++ source code~\cite{Kusano2013}. Delgado-Perez et al. have expanded on the work done for the C language by adding class mutation operators, and created a set of C++ mutation operators~\cite{Delgado-Perez2017}. In addition, they show that the class mutation operators compliment the traditional ones and help testers in developing better test suites.
\vspace*{-1em}
\subsection{C++11/14}
C++11 was introduced in 2011 with the goal of adapting C++ and its core libraries to modern use cases of the language (e.g. multi-threading, genetic algorithms, \ldots). This release was followed by C++14 in 2014 with similar goals.  The introduction of C++11/14 has changed the language to the point that earlier iterations of the language are dubbed the classical C++, and modern C++\footnote{\url{http://www.modernescpp.com/index.php/what-is-modern-c}} starts with C++11/14. The release of the standard was followed by real-time adoption in compilers such as Clang and G++.

Unfortunately, the C++11/14 standard also provides a few features that may cause subtle faults, thus where support in the form of new mutation operators would be desirable. In this subsection we briefly explain these features of C++11/14.

\vspace*{-1em}
\subsubsection{Range-Based For Loop} [\href{http://en.cppreference.com/w/cpp/language/range-for}{http://en.cppreference.com/}] is syntactic sugar made to simplify looping over a range of elements. For example, the following two loops are similar:

\begin{adjustbox}{max width=\textwidth}
	\begin{tabular}{p{0.5\textwidth} p{0.5\textwidth}}
		\begin{lstlisting}
for(int i : v) {
std::cout << i << '\n'; }
		\end{lstlisting}
		&
		\begin{lstlisting}
for(int i=0; i<v.size(); i++) {
std::cout << v.at(i) << '\n'; } 
		\end{lstlisting}
	\end{tabular}
\end{adjustbox}

\vspace*{-2em}
\subsubsection{Lambda Expressions}
[\href{http://en.cppreference.com/w/cpp/language/lambda}{http://en.cppreference.com/}] allow for the definition of unnamed in-line functions. For example, in the following piece of code, lambda contains a function which \textit{captures} \texttt{a} and \texttt{b} (they are available in the body of lambda as const expressions), takes an input parameter \texttt{x}, and returns a \texttt{bool}. 

\begin{lstlisting}
int a, b;
auto lambda = [a, b](int x) {return x > a + b;}
\end{lstlisting}

It is possible to have a default capture at the start of the capture list, e.g. '=' for by-value, or  '\&' for by-reference capture. This causes all variables referenced in the lambda body to be captured the specified way.

\vspace*{-1em}
\subsubsection{Move Semantics} [\href{http://en.cppreference.com/w/cpp/language/move_constructor}{http://en.cppreference.com/}] are introduced in C++11/14 to address the inefficiencies of copy construction when the copied value is deleted after the execution of the constructor. For example, the following code would be inefficient in C++03: 
\begin{lstlisting}
std::vector<int> v(ComputeLargeVector(1000));
\end{lstlisting}

In C++03, this code would create the vector in \texttt{ComputeLargeVector}, call the copy constructor for v, which copies all elements into a newly allocated buffer, and then destroys the original. With move semantics, \texttt{v} would simply copy the internal size, capacity, and pointer to the elements in the temporary vector and set the members of the temporary vector to 0.

To enable this, value categories\footnote{\url{http://en.cppreference.com/w/cpp/language/value_category}} got redefined in C++11. Every expression is either an \texttt{lvalue}, an \texttt{xvalue}, or a \texttt{prvalue}. %
The difference between these value categories lies in two properties: whether or not they have identity (i.e. it is possible to determine whether two expressions are the same using  an address), and whether they can be moved from (move semantics can bind to the expression). \texttt{lvalues} and \texttt{xvalues} have identity, while \texttt{xvalues} and \texttt{prvalues} can be moved from. All \texttt{rvalues} can bind to \texttt{rvalue} references, which are denoted by \&\&. For example, the signature of the move constructor of vector is:
\begin{lstlisting}
vector<T> (vector<T>&&);
\end{lstlisting}
It is possible to convert an \texttt{lvalue} to an \texttt{xvalue} through \texttt{std::move}, which casts the object to an \texttt{rvalue} reference type.
\vspace*{-1em}
\subsubsection{Perfect Forwarding}
[\href{http://en.cppreference.com/w/cpp/utility/forward}{http://en.cppreference.com/}] allow for forwarding of input arguments to other functions as-is. For example, the \texttt{emplace} family of functions in the standard containers accept any number of arguments and forward them to the constructor of the element type.
The following template function constructs an object of type T with a given argument:

\begin{lstlisting}
template<typename T, typename Arg>
T construct(Arg&& argument) {
  return T{std::forward<Arg>(argument)};
}
\end{lstlisting}

Because \texttt{Arg} is a template parameter, \texttt{Arg\&\&} is a forwarding reference~\cite{Sutter2014}. %
This means that it will resolve to either an \texttt{lvalue} or an \texttt{rvalue} reference depending on \texttt{argument}. If \texttt{argument} is an \texttt{lvalue}, \texttt{std::forward} is a no-op, and if \texttt{argument} is an \texttt{rvalue} reference, it behaves the same way \texttt{std::move} does.

\vspace*{-1em}
\subsubsection{List Initialization}
[\href{http://en.cppreference.com/w/cpp/language/list_initialization}{http://en.cppreference.com/}] is a new syntax introduced in C++11 that allows the initialization of an object from braced initial values. It expands the ability to construct structs and arrays using braced initializer to all types in C++. For example, the following is a valid syntax for creating and initializing an array of \textit{int}:

\begin{lstlisting}
int b {1,2,3,4,5}; 
\end{lstlisting}

Also, a type with a constructor that takes \texttt{std::initializer\_list} as an argument can be initialized using this new syntax. For example, the following declaration of a \texttt{std::vector} creates a vector of integers with 5 elements:

\begin{lstlisting}
std::vector<int> v{1,2,3,4,5};
\end{lstlisting}

\vspace*{-1em}
\section{Study Design}
\label{section:Design}
In this section, we discuss the design of our study. First, we explain our evaluation criteria, and then we describe the process by which we determine the fault categories and create mutation operators. Finally, we present the details of our data set. 

\subsection{Evaluation Criteria}

\subsubsection{RQ1.} Which categories of C++11/14 faults are most likely to be made by programmers, and what are the corresponding mutation operators?\\

\noindent To evaluate the results of this question, the mutation operator needs to fulfill the following criteria: 

\begin{compactitem}
\item Can the mutation operator simulate a fault from the fault category we identified?

\item Is it reasonable to assume that the software developer can create faulty code similar to the generated fault?  %
\end{compactitem}

\noindent We look at guidelines provided by experts concerning the new standards and the common pitfalls mentioned therein. We search for such patterns and select those that can be reconstructed into a mutation operator.

\subsubsection{RQ2.} To what extent do these mutation operators create valid, non-equivalent mutants?
\vspace*{-1em}

\begin{equation}
\label{formula}
Mutation\ Operator\ Score = 1-\frac{E - D}{T - I - D} 
\end{equation}

{ \scriptsize  \noindent \centering  T = Total Number of Mutants, E = Number of Equivalent Mutants, D = Number of Easily-Detectable Equivalent Mutants, I = Number of Invalid Mutants\\ }

\vspace*{1em}

An effective mutation operator generates valid semantic faults. This means that mutation operators need to generate as few equivalent mutants as possible. We borrow this criterion from Delgado-Perez et al. who used it in their study~\cite{Delgado-Perez2015a}. It is also important for each mutant to be valid, i.e. the mutated program compiles without errors. To quantify the effectiveness of each mutation operator, we calculate the percentage of equivalent mutants among the valid mutants after filtering the easily-detectable equivalent mutants. The mutation operator score is then calculated by deducting the mentioned percentage from 100\% (see Equation~\ref{formula}). For each mutation operator, we provide methods to filter easily-detectable equivalent mutants.

To see how our operators work in real-life scenarios, we looked at seven open source projects that are using C++11/14 (see Table~\ref{project-stats}).
Our analysis consists of applying  our mutation operators to create all possible mutants. We do this by manually searching for the code patterns that match (using grep). Then, we manually categorize the resulting mutants  into invalid, equivalent, and valid non-equivalent mutants. %
If a mutant did not change the semantics of the program, we classified it as an equivalent mutant. If the operator created a non-compilable program, we classified the mutant as invalid. Otherwise, we considered the mutant as valid non-equivalent.
\vspace*{-1em}
\subsection{Data Set}
In this subsection, we present the details of our data set. Our data set is publicly available in the replication package available at \url{https://www.parsai.net/files/research/ICTSSRepPak.zip}.

In order to find the common fault patterns related to C++11/14, we looked at the authoritative sources of fault patterns such as those suggested by Scott Meyers in his book titled Effective Modern C++~\cite{Meyers2014}, and C++ Core Guidelines by Bjarne Stroustrup~\cite{CppCore}. We also took into account the standard proposal N3853 by Stephan Lavavej~\cite{Lavavej2014} which points out problems with range-based for loop syntax.

\begin{table}[]
\centering
\caption{Project Statistics}
\label{project-stats}
\begin{tabular}{|c|c|c|c|c|c|}
\hline
\multirow{2}{*}{\textbf{Project}} & \multirow{2}{*}{\textbf{Commit}} & \multicolumn{2}{c|}{\textbf{Size(Lines of Code)}} & \textbf{Number of} & \multirow{2}{*}{\textbf{Team Size}} \\ \cline{3-4}
 &  & \textbf{Production} & \textbf{Test} & \textbf{Commits} &  \\ \hline \hline
i-score & c86cd3d & 108K & 3.5K & 5358 & 14 \\ \hline
C++React  & 1f6ddb7 & 11K & 2K & 417 & 1 \\ \hline
EntityX  & 6389b1f & 9K & 1K & 296 & 28 \\ \hline
Antonie  & 59deb0d & 9K & 0.1K & 306 & 2 \\ \hline
Json  & a09193e & 8K & 18K & 1973 & 59 \\ \hline
Corrade  & ff3b351 & 6.5K & 9.1K & 1898 & 10 \\ \hline
termdb  & bd0fb4a & 783 & 153 & 26 & 2 \\ \hline
\end{tabular}
\end{table}

For the evaluation of the mutation operators, we looked at seven open source projects that use C++11/14 (Table \ref{project-stats}). %
These projects range from a small, several hundred lines of code header-only library, to a full application with over 100,000 lines of code with years of active development:\\

\begin{compactitem}
\item \href{https://github.com/OSSIA/i-score/}{i-score} is an interactive intermedia sequencer, built in Qt. 
\item \textit{\href{https://github.com/schlangster/cpp.react}{C++React}} is a C++11 reactive programming library, based on signals and event streams.
\item \textit{\href{https://github.com/alecthomas/entityx}{EntityX}} is an Entity Component System that uses C++11 features.
\item \textit{\href{https://github.com/beaumontlab/antonie}{Antonie}} is a processor of DNA reads, developed at the Bertus Beaumontlab of the Bionanoscience Department of Delft University of Technology.
\item \textit{\href{https://github.com/nlohmann/json}{Json}} is a single-header library for working with Json with modern C++. 
\item \textit{\href{https://github.com/mosra/corrade}{Corrade}} is a C++11/14 utility library, including several container classes, a signal-slot connection library, a unit test framework, a plugin management library and a collection of other small utilities.
\item \textit{\href{https://github.com/agauniyal/termdb}{termdb}} is a small C++11 library for parsing command-line arguments.
\end{compactitem}

\vspace*{-1em}
\section{Results}
\label{section:Results}
In this section, we present the results of our research. For each mutation operator, first we give its definition, then we discuss the motivation behind it to answer RQ1, and finally we provide our analysis of the data set to answer RQ2.   %

\vspace*{-1em}
\subsection{FOR}
The range-based ``for" reference removal (FOR) operator finds instances of range-based for loops of the form \texttt{for (T\& elem : range)} or \texttt{for (T\&\& elem : range)}, where T is either \texttt{auto} or a concrete type, and removes the reference qualifier from the range declaration. Table~\ref{stats-for} shows the results for this mutation operator.

\begin{adjustbox}{max width=\textwidth}
\begin{tabular}{p{0.5\textwidth} p{0.5\textwidth}}
\begin{lstlisting}[caption={Original For}]
for(auto& elem : range) { ... } 
\end{lstlisting}
  &
  \begin{lstlisting}[caption={Mutated For}]
for(auto elem : range) { ... } 
\end{lstlisting}
\end{tabular}
\end{adjustbox}

\vspace*{-3em}

\subsubsection{Motivation (RQ1).} FOR operator is based on the possibility of confusion over the default value semantics of the new range-based for loop, whereas previous methods of looping over containers resulted in reference semantics. This was noted previously by Stephan Lavavej~\cite{Lavavej2014}.
In his standard proposal, he lists three problems with the most idiomatic-looking range-based for loop, \texttt{for (auto elem : range)}, namely:

\begin{compactitem}
\item It might not compile - for example, \texttt{unique\_ptr}\footnote{\url{http://en.cppreference.com/w/cpp/memory/unique_ptr}} elements are not copyable. This is problematic both for users who won't understand the resulting compiler errors, and for users writing generic code that'll happily compile until someone instantiates it for movable-only elements.

\item It might misbehave at runtime - for example, \texttt{elem = val;} will modify the copy, but not the original element in the range. Additionally, \texttt{\&elem} will be invalidated after each iteration.

\item It might be inefficient - for example, unnecessarily copying \texttt{std::string}.
\end{compactitem}

From a mutation testing perspective, the second reason is the main motivation to create a mutation operator. In the case of a range-based for loop that modifies the elements of a container in-place, the correct and generic way to write it is \texttt{for (auto\&\& elem : range)}. For all cases except for proxy objects and move-only ranges, \texttt{for (auto\& elem : range)} works as well.

This operator is only a minor syntactic change that is easily overlooked even in code review if  such fault pattern is not actively looked for. Surviving mutants of this type can pinpoint the loops whose side effects on container elements are not tested.

\begin{table}[]
\centering
\caption{Results of FOR Operator}
\label{stats-for}
  \begin{tabular}{| l | r | r | r | r | r |}
  \hline
  \textbf{Project} & \textbf{Total} & \textbf{Invalid} & \textbf{Equivalent} & \textbf{Easily Detectable} & \textbf{Score} \\
  \hline \hline
  i-score & 251 & 101 & 115 & 110 & 87.5\% \\ \hline
  Corrade  & 24 & 1 & 13 & 13 & 100\% \\ \hline
  Json  & 1 & 0 & 0 & 0 & 100\% \\ \hline
  EntityX & 2 & 0 & 2 & 2 & N/A \\ \hline
  termdb  & 0 & 0 & 0 & 0 & N/A \\ \hline
  C++React & 8 & 0 & 6 & 6 & 100\% \\ \hline
  Antonie  & 39 & 10 & 18 & 18 & 100\% \\ \hline
\end{tabular}
\end{table}

\vspace*{-2em}
\subsubsection{Analysis (RQ2).} \textit{Invalid Mutants:} The invalid mutants are comprised of two groups. The majority of the invalid loops were over containers of move-only types. Of the invalid mutants in \textit{i-score}, 33 were containers of pointers to virtual interface classes with custom dereferencing iterators, making the mutant try to instantiate a non-instantiable type. Both of these cases can be easily checked when generating the mutants.

\noindent \textit{Equivalent Mutants:}
In the majority of equivalent cases, the body of the loop did not mutate the referenced element in the container, thus making it equivalent to a loop with an added \texttt{const} qualifier. This is relatively easy to verify automatically, hence such mutants are listed as detectable. Only a handful of equivalent cases were loops that did mutate the elements of the container, but the container never gets used after the loop finishes. This would require more complicated static analysis. %

\subsection{LMB}
The lambda reference capture (LMB) operator changes a default \textit{value} capture to a default \textit{reference} capture. Table~\ref{stats-lmb} shows the results for this mutation operator.

\begin{adjustbox}{max width=\textwidth}
\begin{tabular}{p{0.5\textwidth} p{0.5\textwidth}}
\begin{lstlisting}[caption={Original Lambda}]
[=](int x) { return x + a; };
\end{lstlisting}
  &
  \begin{lstlisting}[caption={Mutated Lambda}]
[&](int x) { return x + a; };
\end{lstlisting}
\end{tabular}
\end{adjustbox}

\vspace*{-3em}
\subsubsection{Motivation (RQ1).}
This operator is based on the warnings on default capture modes in Core Guideline F53 and Meyers' 31st item~\cite{CppCore,Meyers2014}. This mutation operator results in code that leads to undefined behavior if the lambda is executed in a non-local context, because the references to local variables are not valid. This can happen when the lambda is pushed up the call stack or sent to a different thread for asynchronous execution.

Just like the FOR operator, this operator is only a minor syntactical change that can easily be overlooked, and results in faults that are not necessarily easy to detect; thus it is worth testing for its absence.
Mutants created by this operator are not easy to detect either, because they invoke undefined behavior which is highly dependent on compiler optimization levels and runtime circumstances.

\begin{table}[]
\centering
\caption{Results of LMB Operator}
\label{stats-lmb}
  \begin{tabular}{| l | r | r | r | r | r |}
  \hline
  \textbf{Project} & \textbf{Total} & \textbf{Invalid} & \textbf{Equivalent} & \textbf{Easily Detectable} & \textbf{Score} \\
  \hline \hline
  i-score & 189 & 0 & 113 & 101 & 86.3\% \\ \hline
  Corrade  & 0 & 0 & 0 & 0 & N/A \\ \hline
  Json  & 0 & 0 & 0 & 0 & N/A \\ \hline
  EntityX  & 0 & 0 & 0 & 0 & N/A \\ \hline
  termdb  & 0 & 0 & 0 & 0 & N/A \\ \hline
  C++React & 1 & 0 & 0 & 0 & 100\% \\ \hline
  Antonie  & 0 & 0 & 0 & 0 & N/A \\ \hline
\end{tabular}
\end{table}
\vspace*{-2em}

\subsubsection{Analysis (RQ2).}
\textit{Invalid Mutants:}
We did not witness %
any invalid mutants generated by this operator in our data set.

\noindent \textit{Equivalent Mutants:}
All undetectable equivalent mutations were ones where the lambda gets passed into a function that executes it within its own scope. While it is theoretically possible to detect them, %
we classify them as undetectable because it would require complicated non-local reasoning.
The other equivalent mutants are detectable by taking into account what the capture list actually captures. For example, in Code Excerpt~\ref{lst:LMB1}, the minimal capture list is empty, whereas in Code Excerpt~\ref{lst:LMB2} the minimal capture list is \texttt{[a]} and in Code Excerpt~\ref{lst:LMB3} the minimal capture list is \texttt{[this]}. In the first and third examples, replacing the default value-capture with reference-capture changes nothing about the capture list. In i-score, these made up the majority of equivalent cases, hence the high percentage of detectable equivalent mutants.

\begin{lstlisting}[label={lst:LMB1},caption={Empty Capture}]
[=](int x) {return x < 1;};
\end{lstlisting}

\begin{lstlisting}[label={lst:LMB2},caption={Local Capture}]
int a; [=](int x) {return x < a;};
\end{lstlisting}

\begin{lstlisting}[label={lst:LMB3},caption={`this` Capture}]
struct Foo {
  int a;
  auto getFilter() {
    return [=](int x) {return x < a;};
  }
};
\end{lstlisting}

\subsection{FWD}
The forced \texttt{rvalue} forwarding (FWD) operator replaces  \texttt{std::forward} instances with \texttt{std::move} to force moving from forwarded arguments. Table~\ref{stats-fwd} shows the results for this mutation operator.

\begin{adjustbox}{max width=\textwidth}
\begin{tabular}{p{0.5\textwidth} p{0.5\textwidth}}
\begin{lstlisting}[caption={Original Forwarding}]
template<class T>
void wrapper(T&& arg) 
{
    foo(std::forward<T>(arg));
}
\end{lstlisting}
  &
  \begin{lstlisting}[caption={Mutated Forwarding}]
template<class T>
void wrapper(T&& arg)
{
    foo(std::move(arg));
}
\end{lstlisting}
\end{tabular}
\end{adjustbox}

\vspace*{-3em}
\subsubsection{Motivation (RQ1).}
There are often two possible errors in relation to forwarding semantics (which Meyers warns about in his items 24 and 25~\cite{Meyers2014}): forgetting to use \texttt{std::forward} (and thus passing both \texttt{lvalues} and \texttt{rvalues} on as \texttt{lvalues}) or moving instead of forwarding (and thus passing \texttt{lvalues} on as \texttt{rvalues} to be moved from).

As an example, the following function constructs an object of type T using uniform initialization by forwarding the variadic list of arguments using perfect forwarding:

\begin{lstlisting}
template<typename T, typename... Args>
T construct(Args&&... args) {
  return T{std::forward<Args>(args)...};
}
\end{lstlisting}

We then use the following type, chosen because \texttt{std::string} has a destructive move constructor and \texttt{std::unique\_ptr} is a move-only type:

\begin{lstlisting}
struct Widget
{
  std::string text;
  std::unique_ptr<int> value;
};
\end{lstlisting}

Then the following code constructs two Widgets with the same text and different values:

\begin{lstlisting}
std::string text{64,'a'}; //Long enough to disable SSO
auto w1 = construct<Widget>(text,std::make_unique<int>(0));
auto w2 = construct<Widget>(text,std::make_unique<int>(1));
\end{lstlisting}

Both calls result in \texttt{Args} being \texttt{[std::string\&,std::unique\_ptr<int>\&\&]}, which makes \texttt{std::forward} correctly forward the first argument as \texttt{lvalue} and the second as \texttt{rvalue}. Forgetting to use \texttt{std::forward} results in both arguments being forwarded as \texttt{lvalues}, which fails to compile since \texttt{std::unique\_ptr} is a move-only type. When forgetting to forward, code will always either compile and default to copying the types, or fail to compile because a move-only type is used. Since for all types, the only visible effect of doing a copy instead of a move is a performance degradation, this would not be a useful operator for testing purposes.

Replacing the \texttt{std::forward} with \texttt{std::move}, however, does has the potential to change program behavior. With \texttt{construct} mutated as in the code sample above, the string \texttt{text} will be moved from in the first call, and the second call results in unspecified behavior. In most standard library implementations, \texttt{w2} will end up with an empty text. Meyers argues that it is easy to confuse \texttt{rvalue} and forwarding references because of their identical syntax, making this a likely fault for developers to make.

A large part of these mutants can be targeted by using forwarding on a non-const \texttt{lvalue} argument, since it cannot bind to an \texttt{rvalue} reference. %
Another way of testing these is to use a type with a destructive move, and test the state of the original object after passing it into the function as an \texttt{lvalue}.

\begin{table}[]
\centering
\caption{Results of FWD Operator}
\label{stats-fwd}
  \begin{tabular}{| l | r | r | r | r | r |}
  \hline
  \textbf{Project} & \textbf{Total} & \textbf{Invalid} & \textbf{Equivalent} & \textbf{Easily Detectable} & \textbf{Score} \\
  \hline \hline
  i-score & 71 & 13 & 18 & 9 & 81.6\% \\ \hline
  Corrade  & 5 & 0 & 0 & 0 & 100\% \\ \hline
  Json  & 14 & 0 & 14 & 6 & 0\% \\ \hline
  EntityX  & 7 & 0 & 1 & 1 & 100\% \\ \hline
  termdb  & 0 & 0 & 0 & 0 & N/A \\ \hline
  C++React  & 160 & 0 & 17 & 15 & 98.6\% \\ \hline
  Antonie  & 0 & 0 & 0 & 0 & N/A \\ \hline
\end{tabular}
\end{table}

\subsubsection{Analysis (RQ2).}
\textit{Invalid Mutants:}
The invalid mutants were comprised of two groups: fixed template argument and non-const \texttt{lvalue} reference callee arguments.
The first group forwards to another template function while explicitly stating the template argument as seen in Code Excerpt~\ref{lst:fixedFWD}. This causes the code to not compile when called with a non-const \texttt{lvalue}. If it is called with const \texttt{lvalues} or \texttt{rvalue} references it will have the same runtime behavior as the original.

\begin{lstlisting}[caption={Fixed Template Argument Forwarding},label={lst:fixedFWD}]
template<typename T>
void foo(T&&);

template<typename T>
void bar(T&& t) {
  foo<T>(std::forward<T>(t));
}
\end{lstlisting}

The second group forwards into a function with fixed arguments, at least one of which is a non-const \texttt{lvalue} reference, as seen in Code Excerpt~\ref{lst:lvalueFWD} which defines a function that calls another with a prepended integer argument. Because the second argument is a non-const \texttt{lvalue} reference, applying the operator here results in an invalid mutant because it cannot bind to an \texttt{rvalue} reference.

\begin{lstlisting}[caption={Forwarding into Non-Const Lvalue Reference},label={lst:lvalueFWD}]
void foo(int,int&,int);

template<typename... Args>
void bar(Args&&... args) {
  foo(1,std::forward<Args>(args)...);
}
\end{lstlisting}

\noindent \textit{Equivalent Mutants:} There are three categories of equivalence for this operator. The first is where std::forward gets used within a \texttt{decltype} or \texttt{noexcept} context, where the operator either changes nothing, or makes the code fail to compile. This is why we classify these as detectable equivalent mutants.
The second case is where the forwarded argument never gets stored, which makes irrelevant the difference between \texttt{std::forward}, \texttt{std::move}, and passing by reference. %
The third and final category is where the callees are guaranteed to not take \texttt{rvalue} references or value parameters of movable types.
Of these three categories, the first is easily detectable by filtering out mutants within a \texttt{decltype} or \texttt{noexcept} expression. The second would require sophisticated flow analysis which is why we listed them as not easily-detectable.
The last category can be detected if it is feasible to find all possible callees and see whether they take any \texttt{rvalue} references or value parameters of movable types. This is only feasible for mutants calling functions that cannot be overloaded by external code, since it is otherwise theoretically possible to introduce a new overload of the called function that takes a parameter of a type with a destructive move, making the mutant non-equivalent. The mutants for which this analysis is possible are listed as detectable in our analysis.

\subsection{INI}
The initializer list constructor (INI) operator checks constructor calls of types with an initializer list constructor and changes to/from uniform initialization in order to provoke calling a different constructor.
Table~\ref{stats-ini} shows the results for this mutation operator.

\begin{adjustbox}{max width=\textwidth}
\begin{tabular}{p{0.5\textwidth} p{0.5\textwidth}}
\begin{lstlisting}[caption={Original Initializer},label={lst:oINI}]
std::vector<int> v(3,2);
\end{lstlisting}
  &
  \begin{lstlisting}[caption={Mutated Initializer},label={lst:mINI}]
std::vector<int> v{3,2};
\end{lstlisting}
\end{tabular}
\end{adjustbox}

\vspace*{-3em}
\subsubsection{Motivation (RQ1).}
While initializer list constructors are helpful in defining container contents, they are possible sources of faults as well. For example, when using uniform initialization one needs to pay attention to the correct syntax, since using \{\} instead of () by mistake changes the semantics of the expression drastically. A prominent example of this problem is \texttt{std::vector} of integer types, which Meyers points out in his 7th item~\cite{Meyers2014}.
The non-mutated version in Code Excerpt~\ref{lst:oINI} defines a vector of three elements with value 2, whereas the mutated vector in Code Excerpt~\ref{lst:mINI} has only two elements: 3 and 2.

\begin{table}[]
	\centering
	\caption{Results of INI Operator}
	\label{stats-ini}
	\begin{tabular}{| l | r | r | r | r | r |}
		\hline
		\textbf{Project} & \textbf{Total} & \textbf{Invalid} & \textbf{Equivalent} & \textbf{Easily Detectable} & \textbf{Score} \\
		\hline \hline
		i-score & 1 & 0 & 0 & 0 & 100\% \\ \hline
		Corrade & 0 & 0 & 0 & 0 & N/A  \\ \hline
		Json  & 0 & 0 & 0 & 0 & N/A \\ \hline
		EntityX  & 0 & 0 & 0 & 0 & N/A \\ \hline
		termdb  & 1 & 0 & 0 & 0 & 100\% \\ \hline
		C++React  & 0 & 0 & 0 & 0 & N/A \\ \hline
		Antonie  & 18 & 0 & 0 & 0 & 100\% \\ \hline
	\end{tabular}
\end{table}

\vspace*{-1em}
\subsubsection{Analysis (RQ2).} \textit{Invalid Mutants:}
This operator has no way of creating invalid mutants by design, because it checks whether or not a different constructor is called when it is applied. This includes checking for narrowing conversions; e.g. when trying to mutate \texttt{std::vector<char>(10,'a');}.

\begin{lstlisting}[caption={Equivalence Cases for INI},label={lst:eqINI}]
struct Default1 {
int foo = 1;
Default() = default;
Default(int f) : foo(f) {};
};

std::vector<Default1> v1(1); //v1{1}
std::vector<int> v2(2,2); //v2{2,2}
\end{lstlisting}

\noindent\textit{Equivalent Mutants:}
There are only a few corner cases for \texttt{std::vector} where this operator results in equivalence (e.g. Code Excerpt~\ref{lst:eqINI}).

In both of these cases, the mutated initializer results in the same vector as the original. Given the number of times this pattern was observed in our data set (20 instances in all projects), it is unlikely that such equivalent mutants are found in any significant number.

\subsection{Discussion}

\begin{figure}
	\begin{minipage}{0.5\textwidth}

\centering
\includegraphics[width=\linewidth]{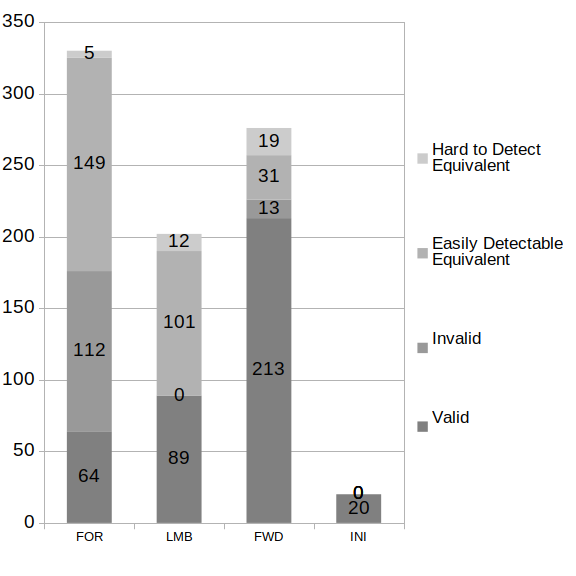}
\caption{Generated Mutants}
\label{figure:summaryMutants}
\end{minipage}	
	\begin{minipage}{0.5\textwidth}
	\centering
	\includegraphics[width=\linewidth]{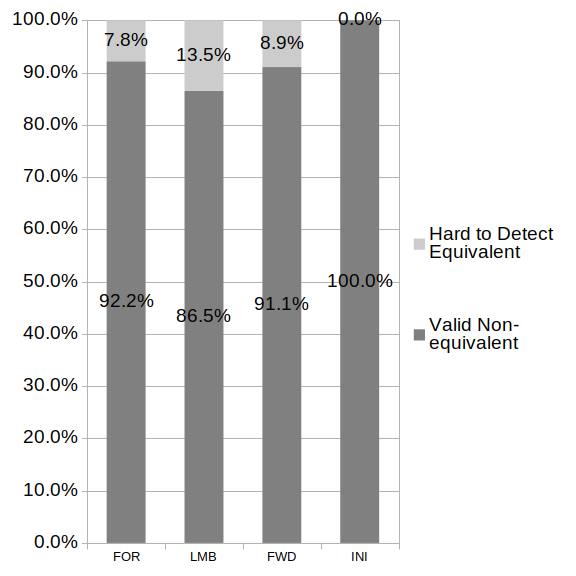}
		\caption{Mutation Operator Scores}
				\label{figure:summaryMOS}
\end{minipage}	
\end{figure}

We have aggregated the number of all generated mutants per kind for each mutation operator in Figure~\ref{figure:summaryMutants}. %
The FOR operator generates the highest number of mutants, most of which are either invalid or easily detectable equivalent. Hence, it is possible to filter most of these mutants easily. This is why this mutation operator is promising. The most promising mutation operator is INI, which generated no invalid or equivalent mutants in our data set. However, the low number of mutants generated by this mutation operator means that it might not be applicable in every case. %
FWD is the operator that generates the most valid, non-equivalent mutants along with a low number of equivalent and invalid mutants, while LMB generates no invalid mutants at all but has a slightly higher ratio of equivalent mutants that are hard to detect.

Figure~\ref{figure:summaryMOS} %
shows the mutation operator score for each mutation operator.
It is clear that all mutation operators are within reasonable boundaries regarding the percentage of generated hard to detect equivalent mutants when compared to other C++ mutation operators (e.g. Delgado-Perez et al.~\cite{Delgado-Perez2015a}).
Overall, we found that these mutation operators have a high mutation operator score, with all of them generating very few equivalent mutants (13.5\% or less of the total number of mutants).

One of the noticeable trends among these mutation operators is their tendency to generate lots of mutants in a single project, and few in others. For example, INI generated 18 mutants in Antonie, and 2 in all other projects, while LMB generated 189 mutants in i-score and only 1 in others. Other than the size of the projects, we found that the adoption of the new syntax has not been uniform in  all of the projects, i.e. some projects make use of mostly a single new syntactic feature and not all of them.

\section{Conclusions and Future Work}
\label{section:Conclusions}
In this study, we created a set of mutation operators that target the common faults introduced by  C++11/14 syntactic features. 
We collected advice about the new C++11/14 syntax from authoritative sources, and created four new statement-level mutation operators (FOR, LMB, FWD, and INI).
For each mutation operator, we discussed the motivation behind its creation and the type of faults they  generate. 
We used Mutation Operator Score as a way to measure the effectiveness of each mutation operator. For this, we selected 7 real-life C++11/14 projects, and counted the number of valid, invalid, easily detectable and hard to detect equivalent mutants generated by each mutation operator for each project. Our results show that all of the introduced mutation operators generate at most 13.5\%  hard to detect equivalent mutants.
The high operator scores indicate that these mutation operators are a useful addition to the mutation operators suggested previously in literature.

Several aspects of this study can be researched further. In particular, the use of our proposed mutation operators alongside traditional and class mutation operators may result in finding multiple redundancies among these mutation operators. In addition, a comparative study similar to Delgado-Perez et al.~\cite{Delgado-Perez2017} between these mutation operator sets would provide more insight into the usefulness of each set of operators depending on the context. %

\vspace*{2em}
{\noindent \small \textbf{Acknowledgments.} This work is sponsored by
(a) the ITEA3 \textsf{ReVaMP$^2$ Project} (number 15010), sponsored by VLAIO---Flanders Innovation Sponsoring Agency;
(b) Flanders Make vzw, the strategic research centre for the manufacturing industry.
}

\bibliographystyle{splncs04}
\bibliography{references}

\end{document}